\documentclass[aps,prl,reprint,twocolumn,superscriptaddress,floatfix,nofootinbib]{revtex4-1}
\usepackage{epsfig,amsmath,amssymb,color,comment,physics}
\usepackage[makeroom]{cancel}
\usepackage[caption=false]{subfig}
\usepackage{mathrsfs}
\usepackage[countmax]{subfloat}
\usepackage[normalem]{ulem}
\usepackage[english]{babel}
\usepackage{dsfont}
\usepackage[bookmarks=true,colorlinks,linkcolor=OrangeRed,urlcolor=NavyBlue,citecolor=RoyalBlue]{hyperref}
\usepackage[dvipsnames]{xcolor}

\newcommand{\Fig}[1]{Fig.~\ref{#1}}

\usepackage{color}

\definecolor{mygold}{rgb}{0.93,0.49,0.13}

\definecolor{blue}{rgb}{0.3,0.3,1.2}

\definecolor{darkgreen}{rgb}{0,0.5,0}

\definecolor{mygreen}{rgb}{0,0.5,0}

\definecolor{purple}{rgb}{0.7,0.3,0.65}

\begin{document}

\title{Thermalization dynamics of a gauge theory on a quantum simulator}

\author{Zhao-Yu Zhou}
\thanks{Z.-Y.Z.~and G.-X.S.~contributed equally to this work.}
\affiliation{Hefei National Research Center for Physical Sciences at Microscale, University of Science and Technology of China, Hefei, China}
\affiliation{School of Physics, University of Science and Technology of China, Hefei, China}
\affiliation{Physikalisches Institut, Ruprecht-Karls-Universit\"at Heidelberg, Heidelberg, Germany}
\affiliation{CAS Centre for Excellence and Synergetic Innovation Centre in Quantum Information and Quantum Physics, University of Science and Technology of China, Hefei, China}

\author{Guo-Xian Su}
\thanks{Z.-Y.Z.~and G.-X.S.~contributed equally to this work.}
\affiliation{Hefei National Research Center for Physical Sciences at Microscale, University of Science and Technology of China, Hefei, China}
\affiliation{School of Physics, University of Science and Technology of China, Hefei, China}
\affiliation{Physikalisches Institut, Ruprecht-Karls-Universit\"at Heidelberg, Heidelberg, Germany}
\affiliation{CAS Centre for Excellence and Synergetic Innovation Centre in Quantum Information and Quantum Physics, University of Science and Technology of China, Hefei, China}

\author{Jad C.~Halimeh}
\affiliation{INO-CNR BEC Center and Department of Physics, University of Trento, Trento, Italy}

\author{Robert Ott}
\affiliation{Institute for Theoretical Physics, Ruprecht-Karls-Universität Heidelberg, Heidelberg, Germany}

\author{Hui Sun}
\affiliation{Hefei National Research Center for Physical Sciences at Microscale, University of Science and Technology of China, Hefei, China}
\affiliation{School of Physics, University of Science and Technology of China, Hefei, China}
\affiliation{Physikalisches Institut, Ruprecht-Karls-Universit\"at Heidelberg, Heidelberg, Germany}
\affiliation{CAS Centre for Excellence and Synergetic Innovation Centre in Quantum Information and Quantum Physics, University of Science and Technology of China, Hefei, China}

\author{Philipp Hauke}
\affiliation{INO-CNR BEC Center and Department of Physics, University of Trento, Trento, Italy}

\author{Bing Yang}
\thanks{Current Address: Department of Physics, Southern University of Science and Technology, Shenzhen 518055, China}
\affiliation{Physikalisches Institut, Ruprecht-Karls-Universit\"at Heidelberg, Heidelberg, Germany}
\affiliation{Institut f\"ur Experimentalphysik, Universit\"at Innsbruck, Innsbruck, Austria}

\author{Zhen-Sheng Yuan}
\affiliation{Hefei National Research Center for Physical Sciences at Microscale, University of Science and Technology of China, Hefei, China}
\affiliation{School of Physics, University of Science and Technology of China, Hefei, China}
\affiliation{Physikalisches Institut, Ruprecht-Karls-Universit\"at Heidelberg, Heidelberg, Germany}
\affiliation{CAS Centre for Excellence and Synergetic Innovation Centre in Quantum Information and Quantum Physics, University of Science and Technology of China, Hefei, China}
\affiliation{Hefei National Laboratory, Hefei, China}

\author{J\"urgen Berges}
\affiliation{Institute for Theoretical Physics, Ruprecht-Karls-Universität Heidelberg, Heidelberg, Germany}

\author{Jian-Wei Pan}
\affiliation{Hefei National Research Center for Physical Sciences at Microscale, University of Science and Technology of China, Hefei, China}
\affiliation{School of Physics, University of Science and Technology of China, Hefei, China}
\affiliation{Physikalisches Institut, Ruprecht-Karls-Universit\"at Heidelberg, Heidelberg, Germany}
\affiliation{CAS Centre for Excellence and Synergetic Innovation Centre in Quantum Information and Quantum Physics, University of Science and Technology of China, Hefei, China}
\affiliation{Hefei National Laboratory, Hefei, China}

\begin{abstract}
    Gauge theories form the foundation of modern physics, with applications ranging from elementary particle physics and early-universe cosmology to condensed matter systems. We perform quantum simulations of the unitary dynamics of a U(1) symmetric gauge field theory and demonstrate emergent irreversible behavior. The highly constrained gauge theory dynamics is encoded in a one-dimensional Bose--Hubbard simulator, which couples fermionic matter fields through dynamical gauge fields. We investigate global quantum quenches and the equilibration to a steady state well approximated by a thermal ensemble. Our work may enable the investigation of elusive phenomena, such as Schwinger pair production and string-breaking, and paves the way for simulating more complex higher-dimensional gauge theories on quantum synthetic matter devices.
\end{abstract}

\date{\today}
\maketitle

Gauge theories provide a fundamental description of quantum dynamics in the Standard Model of particle physics. However, unitary quantum evolution admits no loss of information on a fundamental level; thus, describing from first principles the emergence of phenomena such as thermalization in gauge theories is an outstanding challenge in physics. No general method exists that can simulate the time evolution of the underlying complex quantum many-body theory on classical computers \cite{banuls2020simulating,kasper2014fermion}. Much progress on emergent phenomena has been achieved for simpler systems~\cite{berges2001thermalization,rigol2008thermalization}. For gauge fields, however, the direct connection of far-from-equilibrium behavior at early evolution times with the possible late-time approach to thermal equilibrium, as for instance indicated in collisions of heavy nuclei, remains elusive~\cite{berges2020thermalization}.

Quantum simulators open up a way forward to address this long-standing question. In recent years, there has been much progress in the engineering of gauge theories using various quantum resources such as trapped ions \cite{Martinez2016,Kokail2019}, cold atomic gases \cite{Yang2020,Mil2020,Schweizer2019,Gorg2019,Dai2017}, arrays of Rydberg atoms \cite{bernien2017probing,Surace2019a}, and superconducting qubits \cite{klco2018quantum,dejong2021quantum}. Such table-top platforms can give access to a plethora of observables with high resolution in time and space. However, the simulation requires a large-scale system to incorporate the many degrees of freedom required for the complex quantum field dynamics. In addition, because gauge theories are governed by local symmetries, the engineering of the many gauge constraints at each point in space and time during a nonequilibrium evolution provides a major challenge.

Here we perform quantum simulations of the far-from-equilibrium dynamics of a U(1) symmetric gauge field theory and demonstrate the emergence of thermal equilibrium properties at late times. To achieve this, we utilize a large-scale Bose--Hubbard quantum simulator~\cite{Yang2020} and precisely control the highly excited states relevant for the nonequilibrium dynamics of the gauge theory. The system couples fermionic matter through dynamical gauge fields in one spatial dimension, and employs a discrete ``quantum-link''~\cite{Chandrasekharan1997} representation, discussed also in condensed matter physics \cite{Levin2005, Hermele2004} as well as in the context of particle physics \cite{Kogut1983, Brower1999}.

\begin{figure}[h!]
	\flushleft
	\includegraphics[width=85mm]{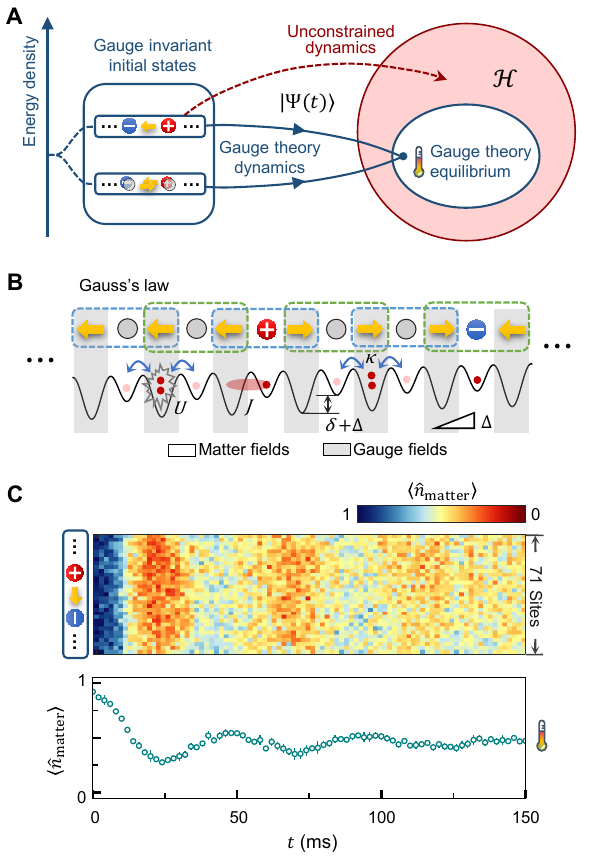}
	\caption{\textbf{Quantum simulation of gauge-theory quench dynamics}. \textbf{(A)}~Schematic nonequilibrium evolution to the steady state. Under the constrained (gauge) condition, we find that different initial states with the same energy density evolve towards a common thermal state of the gauge theory.
	\textbf{(B)}~Quantum simulator for the gauge theory. Matter and gauge fields are represented by occupations of bosonic atoms in an optical superlattice. Charges are illustrated as red (positive) and blue (negative) circles and electric flux is shown as yellow arrows. On matter sites, the presence of an atom signals a corresponding charge in the gauge theory. To illustrate Gauss's law, we indicate locally gauge-invariant configurations around even (green boxes) and odd matter sites (blue boxes), see also Fig.~S1.
	\textbf{(C)}~Evolution of the matter density measured by in-situ imaging. Top: Starting from the initial state with unity-filled matter sites $\langle \hat{n}_\text{matter}\rangle = 1$ (see inset), we observe a fast decay of the matter density $\langle \hat{n}_\text{matter}\rangle$ for ``violent'' quenches ($m/\kappa=0$) in our 71-site quantum simulator. Bottom: Evolution of matter density (averaged over 36 matter sites of the superlattice). Error bars denote the standard deviations. 
	}\label{fmap}
\end{figure}
\begin{figure*}
	\centering
	\includegraphics[width=170mm]{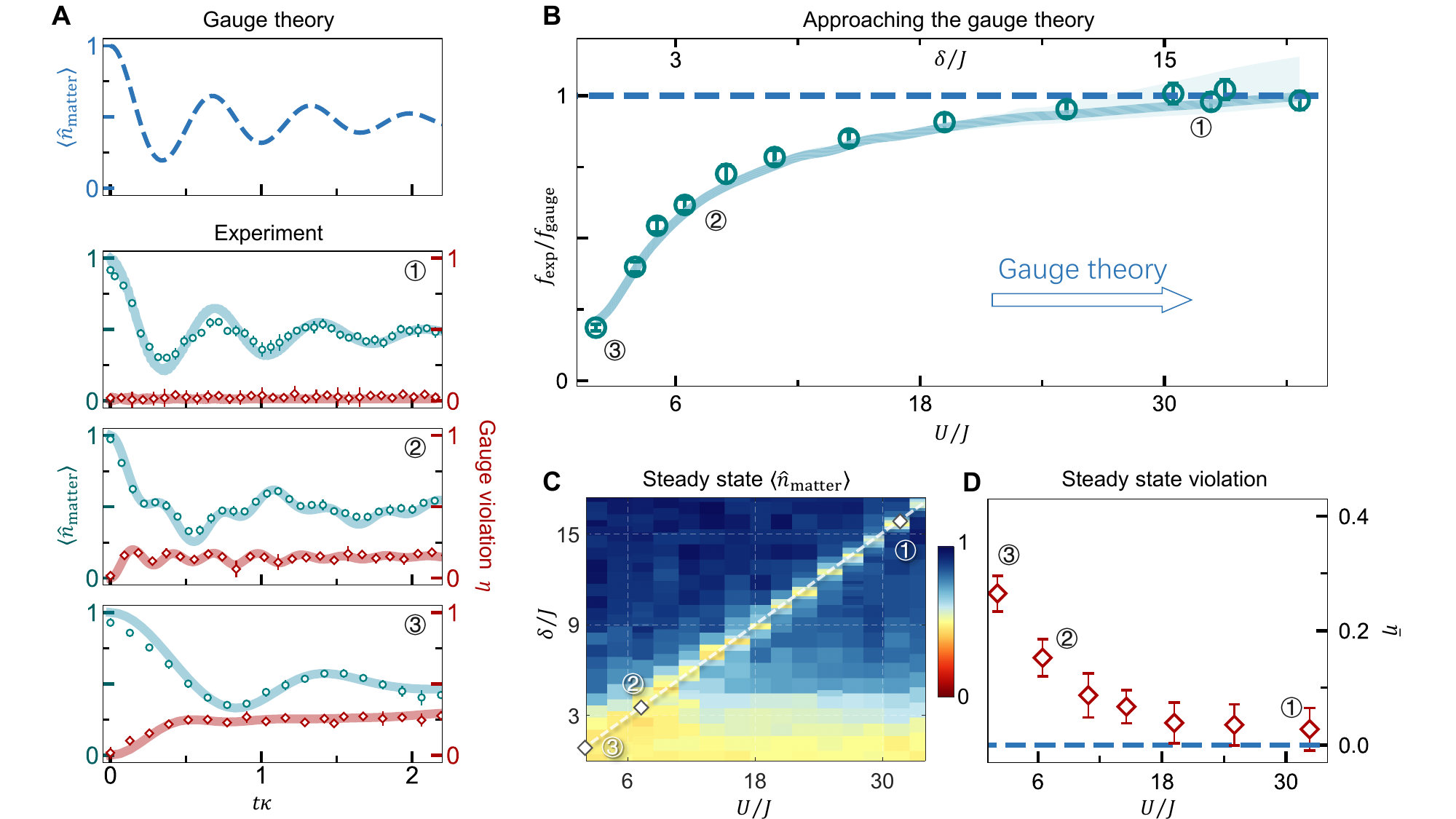}
	\caption{\textbf{Controlled approach to gauge-theory dynamics}. \textbf{(A)} Time evolution in the quantum simulator with and without gauge-theory constraint. The experiment describes the ideal gauge-theory dynamics (top panel) well for case 1 ($U,\delta\gg J$), whereas cases 2 and 3 deviate from these dynamics. The gauge violation, defined as a forbidden odd occupation of gauge sites, decreases from case 3 to case 1 and remains bounded throughout the tractable evolution dynamics. Solid curves are numerical $t$-DMRG results~\cite{supp}. \textbf{(B)} Frequency of many-body oscillations. The experiment ($f_\text{exp}$) shows a fast approach towards the gauge theory ($f_\text{gauge}$) for increasing staggering strength $\delta$. The oscillation frequency is extracted as the lower frequency of a fitted dual-frequency damped sine function (see~\cite{supp}~for details). \textbf{(C)} 2D parameter space. The late-time~($t=120$ ms) matter density is independently scanned over two Hubbard parameters. The dashed white line indicates the resonance condition~($m=0 \leftrightarrow U=2\delta$), applied to all the quench evolution data in the figure. \textbf{(D)} Steady-state gauge violation. The time-averaged gauge violation $\bar\eta$ falls off towards zero for the strongly constrained system with $U/J\gg 1$.
	}
	\label{ffreq}
\end{figure*}

\begin{figure}[h!]
	\centering
	\includegraphics[width=85mm]{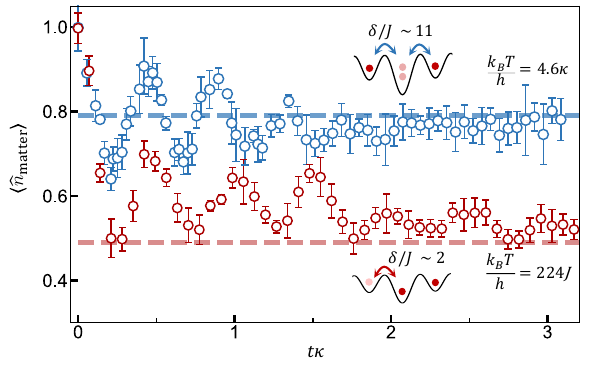}
	\caption{\textbf{Thermalization dynamics with and without gauge-symmetry constraint}. We investigate the real-time evolution of two data sets towards a late-time steady state for constrained (blue) and unconstrained (red) dynamics. Dashed lines show exact-diagonalization predictions from canonical thermal ensembles for the gauge theory (blue) and the Bose--Hubbard model (red) with the same energy density~\cite{supp}. Insets show relevant processes with (top) and without (bottom) the gauge-symmetry constraint.
	}\label{fthermal}
\end{figure}

Exploiting the full experimental tunability of the Bose--Hubbard model parameters, we explore the influence of the gauge-symmetry constraints on the evolution and establish the thermalization dynamics of the $\mathrm{U}(1)$ gauge theory. Following quenching from gauge-invariant initial states far from equilibrium, we observe emergent many-body oscillations through the dynamical annihilation and creation of fermion pairs. We demonstrate an effective loss of information about the system's initial state by starting from different initial conditions with the same conserved quantities and observing relaxation towards a common steady state at longer times. This thermalization dynamics is illustrated in \Fig{fmap}A. 

The unitary dynamics of the one-dimensional lattice gauge theory is governed by the Hamiltonian
\begin{align}
\label{eq:QLM}
\hat{H}_{\text{gauge}}=\sum_{l}\left[\frac{\kappa}{2} \big(\hat{\psi}_{l} \hat{S}_{l,l+1}^+ \hat{\psi}_{l+1} + \text{H.c.}\big) + m \hat{\psi}_l^\dagger \hat{\psi}_l \right],
\end{align}
where $\hat{\psi}_l^{(\dagger)}$ are fermionic field operators~\cite{kogut1975hamiltonian} on matter site $l$ with mass $m$. $\hat{S}_{l,l+1}^{+(-)}$ are spin-$1/2$ raising (lowering) operators for the gauge fields on the link between matter sites $l$ and $l+1$. 
The interaction $\sim \kappa$ represents the annihilation (or creation) of a pair of fermionic charges on neighboring sites with a concomitant change of electric flux $\hat{E}_{l,l+1}=(-1)^{l+1}\hat{S}^z_{l,l+1}$ on the gauge link in-between, such that gauge invariance is retained. The model is realized within a subspace of our quantum simulator, which is described by a tilted Bose--Hubbard Hamiltonian with a staggered potential; see Eq.~S5 for details. It is characterized by direct tunneling strength $J$, staggering potential parameter $\delta$, linear potential $\Delta$, and on-site interaction $U$, as indicated in~\Fig{fmap}B. We employ a Jordan--Wigner transformation to replace the fermionic fields in Eq.~1 with bosonic atoms (see~\cite{supp}~for derivational details).

We keep matter and gauge fields as dynamical degrees of freedom each represented by appropriate site occupations of atoms in an optical superlattice. Gauge symmetry is enforced by suitable energy penalties constraining the system to a gauge-invariant subspace of the quantum simulator~\cite{Zohar2013gauge,halimeh2020reliability,halimeh2020gauge}. For $J\ll \delta, U$, and a linear potential~$\Delta =57$Hz we suppress both direct and long-range tunneling and realize the gauge theory at second-order in perturbation theory~\cite{supp}. We identify the gauge-invariant interaction with a correlated annihilation of two atoms on neighboring matter sites to form a doublon on the gauge link in between (and reverse), see \Fig{fmap}B. The mass of the fermion pair is set by the energy balance of this process as $2m=2\delta-U$ and the interaction strength is given by $\kappa \approx 8\sqrt{2}J^2/U$ close to resonance ($m\sim 0$).

To describe the nonequilibrium evolution of a gauge theory, it is essential to also respect the gauge symmetry in the initial state. In \Fig{fmap}, we show examples of such initial states, which can be prepared in the present apparatus~\cite{Yang2020}. We start the experiment with an array of $36$ near unity-filling chains of $^{87}$Rb atoms in the hyperfine state $\ket{F=1,m_F=-1}$. The individual chains extend over $71$ sites of an optical superlattice, which is formed by the superposition of a short lattice (spacing $a_s=383.5$ nm) and a long lattice (spacing $a_l = 767$ nm). Employing the full tunability of superlattice configurations and the recently developed spin-dependent addressing technique~\cite{Yang2017}, we remove all atoms on odd (gauge) sites, rendering only the even (matter) sites singly occupied in the initial state. The resulting state corresponds to the ground state of Eq.~1 for $\kappa = 0$ and $m<0$, and is characterized by empty gauge sites and unity filling on the matter sites $\langle\hat{n}_\text{matter}\rangle = 1$, where $\langle \hat{n}_\text{matter}\rangle = \sum_{j \in m}\langle \hat{n}_j \rangle/L_\text{m}$ is the average number of bosonic atoms over the $L_\text{m}$ even sites.

After the initial-state preparation, the atoms are isolated in deep lattice wells ($J,\kappa\approx 0$). To initiate the dynamics, we first tune the superlattice configuration such that potential minima of the two lattices are aligned, creating the staggered potential. The quench is then initiated by tuning the laser intensities to realize the desired values of $\kappa$ and $m$, which can be chosen from a broad range.
Subsequently, the system undergoes coherent many-body oscillations. After a certain evolution time, we rapidly ramp up the lattice depth along the $x$-axis to $60E_r$ within $0.1$ ms to freeze the dynamics, where $E_r=h^2/(8m_\text{Rb}a_s^2)$ is the recoil energy with $m_\text{Rb}$ the atom mass and $h$ Planck's constant. We then employ the same site-selective addressing technique and read out $\langle\hat{n}_\text{matter}\rangle$ with in-situ absorption imaging. Each data point is measured by averaging over $6$ realizations of the experiment. We show corresponding in-situ experimental data in \Fig{fmap}C for evolution times $t\leq150$ ms, with $\kappa =14.5$ Hz and $m=0$. For a broad range of model parameters, we observe that the system relaxes towards a steady state after only a few oscillations. The oscillation frequency is mildly affected by the inhomogeneous Gaussian profile of the optical trap towards the edges ($\Delta U\sim 10$ Hz). Overall, the system retains a high degree of homogeneity throughout the tractable evolution times, as demonstrated in~\Fig{fmap}C.

In \Fig{ffreq}, we show the system evolution for the most ``violent'' quench to $m/\kappa=0$, corresponding to $U\approx2\delta$. 
In \Fig{ffreq}A, the real-time dynamics at various microscopic Bose--Hubbard parameters, which all map to the same $m/\kappa$ but with different strengths of the gauge constraint (cases 1-3), is 
compared to theoretical estimates. The top panel shows a result for the ideal gauge field dynamics obtained through exact diagonalization of the Hamiltonian in Eq.~1 for a smaller system with $18$ matter sites. The lower panels give the experimental results for the observable along with numerical estimates based on the time-dependent density matrix renormalization group ($t$-DMRG) \cite{Schollwoeck_review,mptoolkit} for a Bose--Hubbard chain of 32 sites, which show good agreement. In the gauge-theory regime (case 1), we employ the damped-sine fitting at later times~\cite{supp} to extract the damping rate $\gamma$, which is found to be $\gamma^{-1} = 63$ $\pm$ $9$ ms (experiment) and $64.4$ $\pm$ $0.4$ ms ($t$-DMRG). Earliest times can be sensitive to small differences in initial conditions.

The different levels of constrained dynamics are realized by tuning the Bose--Hubbard parameters from $\delta/J = 1$ (case 3) to $\delta/J = 16$ (case 1). This is reflected in the gauge violation~$\eta$, which tends to zero in the gauge-theory regime (case 1). It is defined as the odd atom number expectation value on gauge sites, $\eta = \sum_{j\in g} \langle \hat{n}_{j}\, \mathrm{mod}\, 2 \rangle/L_\text{g}$, where $L_\text{g}$ is the number of odd (gauge) sites~\cite{supp}.
We measure this probability by removing pairs of atoms in the same well with a photo-association laser, followed by selectively addressing the gauge links for imaging.
We use $\eta$ as a measure to validate our quantum simulation of the gauge theory, finding a controlled decrease from large violations in case 3 towards $\eta \approx 0$ in case 1 (Fig.~2, A-D). To characterize the dynamics as a function of the staggering parameter~$\delta/J$, which is used to enforce the gauge constraint, we extract the oscillation frequency of the matter density with a damped-sine fit. In Fig.~2B, our results show a fast approach towards the gauge theory upon increasing~$\delta/J$.

We further investigate the role of the gauge constraint in the relaxation dynamics of the gauge theory by considering quenches to nonzero values of the mass $m$. This amounts to regions away from the resonance line characterized by $U=2\delta$ (\Fig{ffreq}C), where the annihilation of fermion pairs is strongest. For $m=-0.8\kappa$, the resulting time evolution is displayed in \Fig{fthermal} with both weakly ($\delta/J\sim 2$) and strongly constrained ($\delta/J\sim 11$) dynamics shown. Here and in the following, we focus on a region of interest of 50 chains each with an extent of 50 sites. This mitigates the effects of a slightly inhomogeneous trap. We compare the strongly constrained dynamics with the thermal prediction of the gauge theory, finding agreement within experimental precision at late times. In contrast, the unconstrained system evolves towards a very different state, characterized by a thermal ensemble of the Bose--Hubbard system away from the gauge-theory regime. In Fig.~3, the thermal predictions have been obtained from a numerical evaluation of the corresponding microcanonical and canonical ensembles. We extract the temperatures from the latter as shown in \Fig{fthermal} by fixing their (conserved) energy density to that of the pure initial state~\cite{supp}.

\begin{figure*}[t!]
	\centering
	\includegraphics[width=180mm]{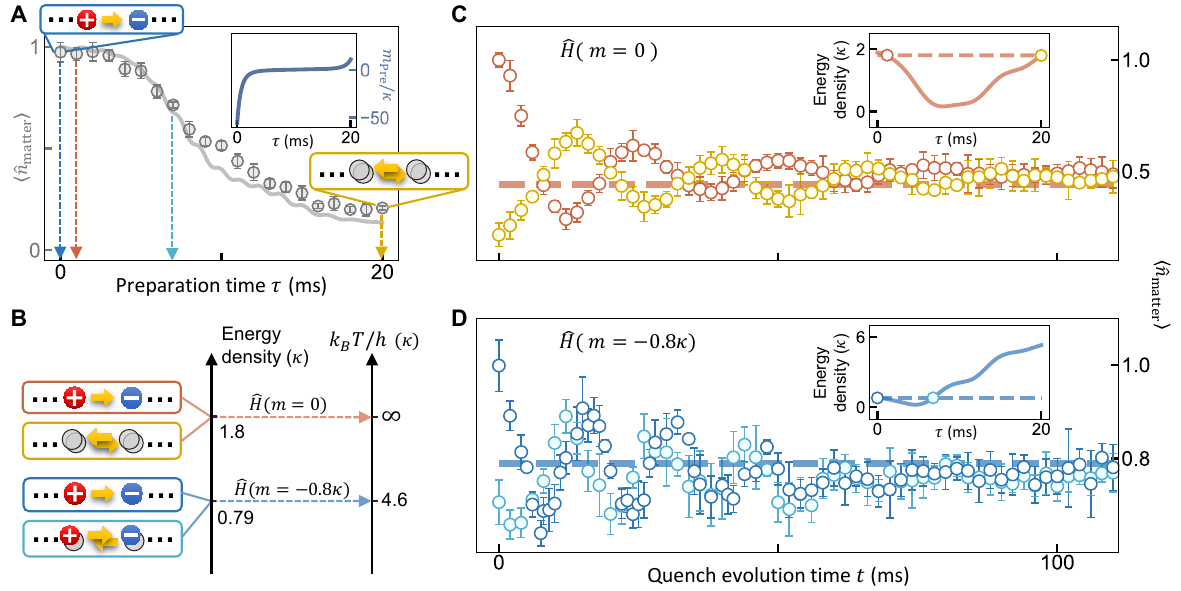}
	\caption{\textbf{Effective loss of initial-state information}. \textbf{(A)} State preparation. Evolution of the matter density from the ``fully matter-filled'' state ($\langle \hat{n}_\text{matter}\rangle = 1$, blue box left) to almost ``matter-empty'' state ($\langle \hat{n}_\text{matter}\rangle \approx 0.21$, yellow box right) for the adiabatic ramp with preparation time $\tau$ and corresponding mass parameter $m_{\mathrm{Pre}}/\kappa$ as shown in the inset. \textbf{(B)} Schematic of the evolution towards thermal equilibrium. For each of two sets of quench parameters ($m=0$ and $m=-0.8\kappa$) we choose two initial states with equal energy density. The resulting steady states in the wake of the quenches starting in these two initial states are then compared to a canonical thermal ensemble whose temperature is determined from the energy density~\cite{supp}. Here, all energy densities are plotted with respect to the ground state of the evolution Hamiltonian. \textbf{(C, D)} Relaxation. We show the thermalization dynamics for the chosen quench parameters and initial states (shown in (B)). Experimental data are compared to predictions from corresponding gauge theory thermal ensembles (dashed lines) at temperatures $k_BT=\infty$ (top) and $k_BT=4.6\kappa$ (bottom). The insets show the energy density evolution during state preparation, the circles mark the chosen initial states.  }\label{finitialcond}
\end{figure*}

As a next step, we investigate the role of the initial state in the thermalization dynamics of the gauge theory. If the system approaches thermal equilibrium, then the late-time behavior is entirely characterized by conserved quantities. The tunability of the Hubbard parameters allows us to access a broad range of gauge-symmetric initial states with an adiabatic ramp, ranging from the fully filled state ($\langle\hat{n}_\text{matter}\rangle = 1$) to states where a large fraction fermion pairs have annihilated ($\langle \hat{n}_\text{matter}\rangle \approx 0.21$)~\cite{Yang2020}, see \Fig{finitialcond}A. For the quench dynamics, we compare initial states with the same energy density with respect to the quench Hamiltonian(\Fig{finitialcond}B). To achieve this, we numerically follow the experimental sequence and determine suitable ramp times as outlined Fig.~4A for two values of $m/\kappa=-0.8$ and $m/\kappa=0$~\cite{supp}. For $m/\kappa=0$, the initial states with the same energy density are prepared with the ramp times 1.2 ms and 20 ms, corresponding to $\langle\hat{n}_\text{matter}\rangle = 1$ and $\langle \hat{n}_\text{matter}\rangle \approx 0.21$ respectively. For $m/\kappa=-0.8$, the ramp times are 0 ms and 6.8 ms, corresponding to the initial states of $\langle\hat{n}_\text{matter}\rangle = 1$ and $\langle \hat{n}_\text{matter}\rangle \approx 0.71$. The ensuing dynamics is characterized by transient many-body oscillations, where the different initial states relax to a common steady state at long times (Fig.~4, C and D). During the evolution, the information about initially different matter densities is seen to be effectively lost in the quantum many-body system. We again find the long-time steady states to be well-described by gauge-invariant thermal ensembles with the same conserved charges as the initial state. We observe this thermalization dynamics for different values of couplings in the gauge theory. Specifically for $m=0$ our initial states are distributed symmetrically around the center of the energy spectrum~\cite{supp}. In this case one observes a fast effective loss of initial-state memory, and the experiment relaxes to the steady-state value of the infinite-temperature state (\Fig{finitialcond} B and C.)

Effectively irreversible behavior, such as thermalization from an underlying reversible time evolution, emerges in general in nonintegrable models for local observables and typical initial states~\cite{Rigol2008thermalization}. Despite the nonintegrability of the $\mathrm{U}(1)$ quantum link model \cite{turner2018weak,Surace2019a,Zache2021}, certain fine-tuned quenches give rise to weak ergodicity breaking owing to the presence of special eigenstates in the spectrum of the quench Hamiltonian~\cite{turner2018weak,Surace2019a}. This could manifest in persistent oscillations around the thermal-ensemble prediction~\cite{bernien2017probing}. Currently, the level of control in our experiment limits us from probing such possible behavior in part because of an inherent residual inhomogeneity across the lattice. Thus, in our experiment we observe equilibration to close to the thermal equilibrium value for all quenches we performed.

Our work creates a pathway for addressing emergent dynamical phenomena in gauge theories, such as the Schwinger effect \cite{zache2018quantum}, dynamical topological quantum phase transitions \cite{Huang2019}, and string-breaking \cite{Banerjee2012,hebenstreit2013real} in strong fields \cite{Narozhny2014}. The approach lays also the foundations for the exploration of more complex higher-dimensional gauge theories using state-of-the-art quantum technology \cite{zohar2021quantum}. An important next step towards applications for gauge theories such as quantum electrodynamics, or maybe even quantum chromodynamics, is a faithful extension of the discrete quantum-link representation towards continuous variables~\cite{Mil2020,prufer2020experimental,Zache:2019xkx}. To this end, current implementation schemes should be extended to higher spin representations and scalable higher-dimensional set-ups~\cite{ott2020scalable}.


\begin{thebibliography}{0}%
\makeatletter
\providecommand \@ifxundefined [1]{%
 \@ifx{#1\undefined}
}%
\providecommand \@ifnum [1]{%
 \ifnum #1\expandafter \@firstoftwo
 \else \expandafter \@secondoftwo
 \fi
}%
\providecommand \@ifx [1]{%
 \ifx #1\expandafter \@firstoftwo
 \else \expandafter \@secondoftwo
 \fi
}%
\providecommand \natexlab [1]{#1}%
\providecommand \enquote  [1]{``#1''}%
\providecommand \bibnamefont  [1]{#1}%
\providecommand \bibfnamefont [1]{#1}%
\providecommand \citenamefont [1]{#1}%
\providecommand \href@noop [0]{\@secondoftwo}%
\providecommand \href [0]{\begingroup \@sanitize@url \@href}%
\providecommand \@href[1]{\@@startlink{#1}\@@href}%
\providecommand \@@href[1]{\endgroup#1\@@endlink}%
\providecommand \@sanitize@url [0]{\catcode `\\12\catcode `\$12\catcode
  `\&12\catcode `\#12\catcode `\^12\catcode `\_12\catcode `\%12\relax}%
\providecommand \@@startlink[1]{}%
\providecommand \@@endlink[0]{}%
\providecommand \url  [0]{\begingroup\@sanitize@url \@url }%
\providecommand \@url [1]{\endgroup\@href {#1}{\urlprefix }}%
\providecommand \urlprefix  [0]{URL }%
\providecommand \Eprint [0]{\href }%
\providecommand \doibase [0]{http://dx.doi.org/}%
\providecommand \selectlanguage [0]{\@gobble}%
\providecommand \bibinfo  [0]{\@secondoftwo}%
\providecommand \bibfield  [0]{\@secondoftwo}%
\providecommand \translation [1]{[#1]}%
\providecommand \BibitemOpen [0]{}%
\providecommand \bibitemStop [0]{}%
\providecommand \bibitemNoStop [0]{.\EOS\space}%
\providecommand \EOS [0]{\spacefactor3000\relax}%
\providecommand \BibitemShut  [1]{\csname bibitem#1\endcsname}%
\let\auto@bib@innerbib\@empty
\end{thebibliography}%


\begin{thebibliography}{10}

\bibitem{banuls2020simulating}
M.~C. Bañuls, R.~Blatt, J.~Catani, A.~Celi, et~al., {\it The European Physical
  Journal D\/} {\bf 74} (2020).

\bibitem{kasper2014fermion}
V.~Kasper, F.~Hebenstreit, J.~Berges, {\it Phys. Rev. D\/} {\bf 90}, 025016
  (2014).

\bibitem{berges2001thermalization}
J.~Berges, J.~Cox, {\it Physics Letters B\/} {\bf 517}, 369 (2001).

\bibitem{rigol2008thermalization}
M.~Rigol, V.~Dunjko, M.~Olshanii, {\it Nature\/} {\bf 452}, 854–858 (2008).

\bibitem{berges2020thermalization}
J.~Berges, M.~P. Heller, A.~Mazeliauskas, R.~Venugopalan, {\it Rev. Mod.
  Phys.\/} {\bf 93}, 035003 (2021).

\bibitem{Martinez2016}
E.~A. Martinez, {\it et~al.\/}, {\it Nature\/} {\bf 534}, 516–519 (2016).

\bibitem{Kokail2019}
C.~Kokail, {\it et~al.\/}, {\it Nature\/} {\bf 569}, 355 (2019).

\bibitem{Yang2020}
B.~Yang, {\it et~al.\/}, {\it Nature\/} {\bf 587}, 392–396 (2020).

\bibitem{Mil2020}
A.~Mil, {\it et~al.\/}, {\it Science\/} {\bf 367}, 1128 (2020).

\bibitem{Schweizer2019}
C.~Schweizer, {\it et~al.\/}, {\it Nature Physics\/} {\bf 15}, 1168 (2019).

\bibitem{Gorg2019}
F.~Görg, {\it et~al.\/}, {\it Nature Physics\/} {\bf 15}, 1161–1167 (2019).

\bibitem{Dai2017}
H.-N. Dai, {\it et~al.\/}, {\it Nature Physics\/} {\bf 13}, 1195–1200 (2017).

\bibitem{bernien2017probing}
H.~Bernien, {\it et~al.\/}, {\it Nature\/} {\bf 551}, 579–584 (2017).

\bibitem{Surace2019a}
F.~M. Surace, {\it et~al.\/}, {\it Physical Review X\/} {\bf 10} (2020).

\bibitem{klco2018quantum}
N.~Klco, {\it et~al.\/}, {\it Physical Review A\/} {\bf 98} (2018).

\bibitem{dejong2021quantum}
W.~A. de~Jong, {\it et~al.\/}, Quantum simulation of non-equilibrium dynamics
  and thermalization in the schwinger model (2021).

\bibitem{Chandrasekharan1997}
S.~Chandrasekharan, U.~J. Wiese, {\it Nuclear Physics B\/} {\bf 492}, 455
  (1997).

\bibitem{Levin2005}
M.~Levin, X.~G. Wen., {\it Reviews of Modern Physics\/} {\bf 77}, 871 (2005).

\bibitem{Hermele2004}
M.~Hermele, M.~P.~A. Fisher, L.~Balents, {\it Physical Review B\/} {\bf 69}
  (2004).

\bibitem{Kogut1983}
J.~B. Kogut, {\it Reviews of Modern Physics\/} {\bf 55}, 775 (1983).

\bibitem{Brower1999}
R.~Brower, S.~Chandrasekharan, U.-J. Wiese, {\it Physical Review D\/} {\bf 60}
  (1999).

\bibitem{supp}
See supplementary materials.

\bibitem{kogut1975hamiltonian}
J.~Kogut, L.~Susskind, {\it Phys. Rev. D\/} {\bf 11}, 395 (1975).

\bibitem{Zohar2013gauge}
E.~Zohar, J.~I. Cirac, B.~Reznik, {\it Phys. Rev. Lett.\/} {\bf 110}, 055302
  (2013).

\bibitem{halimeh2020reliability}
J.~C. Halimeh, P.~Hauke, {\it Phys. Rev. Lett.\/} {\bf 125}, 030503 (2020).

\bibitem{halimeh2020gauge}
J.~C. Halimeh, H.~Lang, J.~Mildenberger, Z.~Jiang, P.~Hauke, {\it PRX
  Quantum\/} {\bf 2}, 040311 (2021).

\bibitem{Yang2017}
B.~Yang, {\it et~al.\/}, {\it Physical Review A\/} {\bf 96} (2017).

\bibitem{Schollwoeck_review}
U.~Schollwöck, {\it Annals of Physics\/} {\bf 326}, 96–192 (2011).

\bibitem{mptoolkit}
I.~P.~McCulloch, \textit{Matrix Product Toolkit}. URL:
  https://people.smp.uq.edu.au/IanMcCulloch/mptoolkit/.

\bibitem{Rigol2008thermalization}
M.~Rigol, V.~Dunjko, M.~Olshanii, {\it Nature\/} {\bf 452},
  854–858 (2008).

\bibitem{turner2018weak}
C.~J. Turner, A.~A. Michailidis, D.~A. Abanin, M.~Serbyn, Z.~Papi{\'c}, {\it
  Nature Physics\/} {\bf 14}, 745 (2018).

\bibitem{Zache2021}
T.~V. Zache, M.~V. Damme, J.~C. Halimeh, P.~Hauke, D.~Banerjee  (2021).

\bibitem{zache2018quantum}
T.~V. Zache, {\it et~al.\/}, {\it Quantum Science and Technology\/} {\bf 3},
  034010 (2018).

\bibitem{Huang2019}
Y.-P. Huang, D.~Banerjee, M.~Heyl, {\it Phys. Rev. Lett.\/} {\bf 122}, 250401
  (2019).

\bibitem{Banerjee2012}
D.~Banerjee, {\it et~al.\/}, {\it Physical Review Letters\/} {\bf 109}, 1
  (2012).

\bibitem{hebenstreit2013real}
F.~Hebenstreit, J.~Berges, D.~Gelfand, {\it Phys. Rev. Lett.\/} {\bf 111},
  201601 (2013).

\bibitem{Narozhny2014}
N.~B. Narozhny, A.~M. Fedotov, {\it European Physical Journal: Special
  Topics\/} {\bf 223}, 1083 (2014).

\bibitem{zohar2021quantum}
E.~Zohar, {\it Phil. Trans. R. Soc. A.\/} {\bf 380} (2021).

\bibitem{prufer2020experimental}
M.~Prüfer, {\it et~al.\/}, {\it Nature Physics\/} {\bf 16}, 1012–1016
  (2020).

\bibitem{Zache:2019xkx}
T.~V. Zache, T.~Schweigler, S.~Erne, J.~Schmiedmayer, J.~Berges, {\it Phys.
  Rev. X\/} {\bf 10}, 011020 (2020).

\bibitem{ott2020scalable}
R.~Ott, T.~V. Zache, F.~Jendrzejewski, J.~Berges, {\it Phys. Rev. Lett.\/} {\bf
  127}, 130504 (2021).

\bibitem{zhaoyu2022data}
Z.-Y. Zhou, G.-X. Su, J.~C. Halimeh, R.~Ott, {\it et~al.\/}, Harvard Dataverse, doi.org/10.7910/DVN/XWUHTS.


\end{thebibliography}

~\\

\noindent \textbf{Acknowledgments}\\

\noindent The authors are grateful to D.~Banerjee, A.~Sen, J.-Y.~Desaules, M.~G\"arttner, A.~Hudomal, F.~Jendrzejewski, A.~Lazarides, M.~Oberthaler, Z.~Papi\'c, J.~Schmiedmayer, C.~J.~Turner, and T.~V.~Zache for discussions. \textbf{Fundings:} This work is part of and supported by NNSFC grant 12125409, Innovation Program for Quantum Science and Technology 2021ZD0302000, the Deutsche Forschungsgemeinschaft (DFG, German Research Foundation) - Project-ID 273811115 - SFB 1225 and under Germany’s Excellence Strategy EXC 2181/1-390900948 (the Heidelberg STRUCTURES Excellence Cluster), the Anhui Initiative in Quantum Information Technologies, the Chinese Academy of Sciences, Provincia Autonoma di Trento, the ERC Starting Grant StrEnQTh (project ID 804305), the Google Research Scholar Award ProGauge, and Q@TN — Quantum Science and Technology in Trento. \textbf{Author contributions:} Z.-Y.Z.~and G.-X.S.~contributed equally to this work. Z.-Y.Z., G.-X.S., and H.S.~performed the experiments and analyzed the data; J.C.H.~and R.O.~performed the numerical calculations and comparisons to experimental data; H.S., B.Y., Z.-S.Y., and J.-W.P.~developed relevant experimental techniques; P.H., B.Y., and J.B. initiated early discussions on this topic; J.B., P.H., J.C.H.,~and~R.O.~developed the theory; Z.-S.Y.~and J.-W.P.~conceived and supervised the experimental research; all authors contributed to the writing of the manuscript. \textbf{Competing interests:} The authors declare no competing financial interests. \textbf{Data and materials availability:} All experimental data and code are available in the database~\cite{zhaoyu2022data}.

~\\

\noindent \textbf{Supplementary materials}\\

\noindent Supplementary Text\\
Figs. S1 to S5\\
References(43-44)


\newpage
\onecolumngrid
\vspace*{0.5cm}
\begin{center}
	\textbf{SUPPLEMENTARY MATERIAL}
\end{center}
\vspace*{0.5cm}
\twocolumngrid
\setcounter{equation}{0}
\setcounter{figure}{0}
\makeatletter
\makeatother
\renewcommand{\theequation}{S\arabic{equation}}
\renewcommand{\thefigure}{S\arabic{figure}}

\subsection{Details on the gauge theory Hamiltonian}
In the main text, we have used a particle-hole transformation to express the spin-$1/2$ $\mathrm{U}(1)$ quantum link model (QLM) in a "charge basis" where excitations of the fermion fields on matter sites relate to the presence of electric charges. Here, we outline this particle-hole transformation in detail. We start with the Hamiltonian of the spin-$1/2$ $\mathrm{U}(1)$ QLM with staggered fermions~\cite{Kogut1983}
\begin{align}\label{eq:QED}
	\hat{H}_{\text{gauge}}=\sum_{l}\left[\frac{\kappa}{2} \big(\hat{\psi}_{l}^\dagger \hat{S}_{l,l+1}^+ \hat{\psi}_{l+1} + \text{H.c.}\big) + m (-1)^l\hat{\psi}_l^\dagger \hat{\psi}_l \right].
\end{align}
Here, the generators of the $\mathrm{U}(1)$ gauge symmetry are the Gauss operators $\hat{G}_l=\hat{\psi}_{l}^\dagger \hat{\psi}_{l}+((-1)^{l+1}-1)/2-\hat{S}^z_{l-1,l}+\hat{S}^z_{l,l+1}$, with eigenvalues $g_l$ that locally determine the gauge-invariant sector. In our work, we choose the target sector as $g_l=0,\,\forall l$. The Hamiltonian~\eqref{eq:QED} is invariant under the transformation $\hat{H}=\hat{V}_l^\dagger \hat{H}\hat{V}_l,\,\forall l$ with the unitary operator $\hat{V}_l=\exp(i\alpha_l\hat{G}_l)$, where $\alpha_l$ is a continuous angle $ \in (0,2\pi]$ which may depend on the lattice site index $l$.

\begin{figure}
	\centering
	\includegraphics[width=80mm]{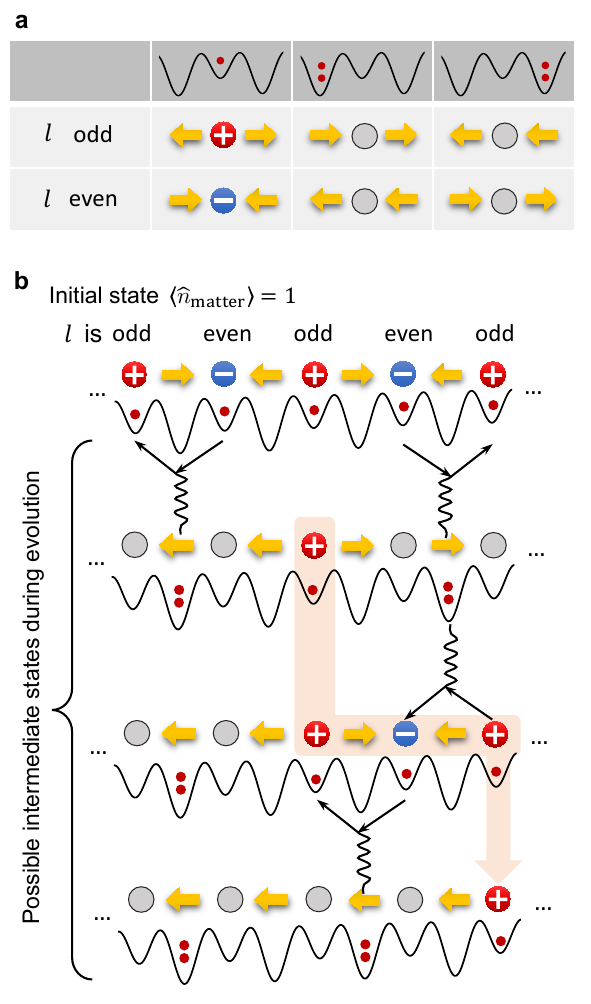}
	\caption{Details on the mapping of the gauge theory. (a)~We display the locally allowed configurations in our quantum simulator and their analogs in the gauge theory according to Gauss law. (b)~We show the frequently used initial state with $\langle \hat{n}_\mathrm{matter}\rangle$ in both gauge theory and quantum simulator and illustrate the transport of fermions in the subsequent dynamics.
	}
	\label{app:Gauss-law}
\end{figure}

To employ the particle-hole transformation, we replace for even matter sites
\begin{subequations}
\begin{align}
&\hat{\psi}^\dagger_{2l}\hat{\psi}_{2l} \to 1-\hat{\psi}^\dagger_{2l}\hat{\psi}_{2l},  \,\,\,\,\,\hat{\psi}_{2l}\to \hat{\psi}^\dagger_{2l},\\
&\hat{S}^z_{2l,2l+1}\to(-1)\hat{S}^z_{2l,2l+1},\,\,\,\,\,\hat{S}^{\pm}_{2l,2l+1}\to\hat{S}^{\mp}_{2l,2l+1}.
\end{align}
\end{subequations}
The Hamiltonian of the spin-$1/2$ $\mathrm{U}(1)$ QLM now becomes Eq.~\eqref{eq:QLM} of the main text, where we additionally drop an irrelevant constant. The generator of the $\mathrm{U}(1)$ gauge symmetry is replaced by
\begin{align}
\hat{G}_l\to&(-1)^{l+1}\big(\hat{S}^z_{l-1,l}+\hat{S}^z_{l,l+1}+\hat{\psi}_{l}^\dagger \hat{\psi}_{l}\big).
\end{align}
Locally, it allows for a set of three states as illustrated in \Fig{app:Gauss-law}\textbf{a}, which in an extended lattice are coupled to gauge-invariant many-body states. The gauge-invariant interaction term can now be readily connected to the creation and annihilation of a pair of electric charges, see \Fig{app:Gauss-law}\textbf{b}. Furthermore, the staggering in the mass term has been removed through the particle-hole transformation.

\subsection{Mapping the Bose--Hubbard simulator onto the $\mathrm{U}(1)$ gauge theory}
The experiment is governed by the Bose--Hubbard Hamiltonian
\begin{align}\label{eq:Hexp}
    \hat{H}_{\mathrm{exp}} = \sum_j \bigg[ J\big(\hat{b}^{\dagger}_j \hat{b}_{j+1}+\text{H.c.}\big)+ \epsilon_j  \hat{n}_{j} +\frac{U}{2}\hat{n}_j\big(\hat{n}_j-1\big)\bigg],
\end{align}
where $\epsilon_j = j\Delta + (-1)^{j}\delta/2$ encodes the potentials from the staggered superlattice and the gravitational gradient, and $\hat{n}_j=\hat{b}^\dagger_j \hat{b}_j$ is the number operator.
For $J\ll \delta, U$ and $U \sim 2\delta$, we derive the effective Hamiltonian (valid in the bosonic occupation subspace defined below)  
\begin{align}
\label{eq:effective-Hamiltonian}
    \hat{H}_{\mathrm{eff}} = \sum_{j\in e} \left[\frac{\kappa}{2\sqrt{2}}\big(\hat{b}_j^\dagger\hat{b}_{j+2}^\dagger \hat{b}_{j+1}^2 + \text{H.c.}\big) + m \hat{b}_j^\dagger\hat{b}_j  \right],
\end{align}
at second order in perturbation theory, where $m = \delta- U/2$, and the sum runs over all even ($j\in e$) optical lattice wells corresponding to the matter sites $l$. The correlated hopping interaction $\kappa$ is given by
\begin{align}
\kappa = 2\sqrt{2}J^2\left[\frac{\delta}{\delta^2-\Delta^2} + \frac{U-\delta}{(U -\delta)^2-\Delta^2}\right].
\end{align}
Here, we have also considered the condition $\Delta \gg \kappa$, to suppress second-order tunneling over two sites. Close to resonance, i.e., $U\approx 2\delta$, and using $\Delta \ll \delta, U$ as realized in our experiment, we have $\kappa \approx 8\sqrt{2}J^2/U$.

For the considered lattice parameters and initial state, the system is effectively constrained to the site occupations $\ket{0}$ and $\ket{1}$ for the even sites of the superlattice (matter sites), and $\ket{0}$ and $\ket{2}$ for the odd sites of the superlattice (gauge links). The Hamiltonian \eqref{eq:effective-Hamiltonian} as well as the following operator identifications are valid in this occupation subsector of the Hilbert space and should be interpreted with the corresponding projection operators $\mathcal{P}_j$ onto the associated local subspace.
To make direct contact to the lattice gauge theory Hamiltonian \eqref{eq:QLM}, we identify the bosonic creation operators on odd superlattice wells ($2l+1=j\in o$) as the gauge fields on the gauge links~$(l,l+1)$
\begin{subequations}
\begin{align}
    \hat{b}^2_{j\in o}  &= \sqrt{2}\hat{S}^{-}_{l,l+1} ,\\
    \hat{n}_{j\in o}&= 2\hat{S}^z_{l,l+1}+1.
\end{align}
\end{subequations}
Further, using a Jordan--Wigner transformation, we map the bosonic creation operators on even superlattice wells, $2l=j\in e$, to fermionic matter field operators $\hat{\psi}_l$
\begin{subequations}
\begin{align}
     \hat{b}_{j\in e}^{(\dagger)} &= (-1)^l e^{(-)i\pi \sum_{l'<l} \hat{\psi}^\dagger_{l'}\hat{\psi}_{l'} }\hat{\psi}_l^{(\dagger)},  \\
     \hat{n}_{j\in e}&= \hat{\psi}_l^\dagger \hat{\psi}_l,
\end{align}
\end{subequations}
These mappings yield the lattice gauge theory Hamiltonian \eqref{eq:QLM}.

\subsection{State preparation and detection}
The experiment begins with a single layer of quasi-2D $\langle \hat{n}\rangle = 1$ Mott insulator of $^{87}$Rb atoms prepared in the hyperfine state $\ket{F=1,m_{F}=-1}$. We employ the staggered-immersion cooling technique with an optical superlattice along the $x$-axis to reach a filling rate of $99.2\%$ \cite{Yang2020}. The optical superlattice consists of two standing waves with laser frequencies $\lambda_{s}=767$ nm and $\lambda_{l}=1534$ nm, forming the potential
\begin{equation}
\label{eq:superlattice}
\ V(x) = V_s \text{cos}^2({k x}) - V_l \text{cos}^2({k x/2}+\varphi),
\end{equation}
where $k = {2\pi}/{\lambda_\text{s}} $ is the wave number of short lattice, and $V_s$ and $V_l$ are the depths of the short and long lattices, respectively. The relative phase $\varphi$ controlled by the relative frequency of these lasers determines the superlattice structure. The quench dynamics is observed at $\varphi = \pi/4$, forming the staggered potential with equal hopping $J$ between neighboring sites, while the site-selective addressing is performed at $\varphi = 0$, indicating a balanced condition for the double well.

To perform the site-selective addressing, we first set the bias magnetic field along the $x$-axis, then by tuning the polarization of the short lattice along the $x$-axis with an electro-optical modulator, the hyperfine transition frequency between the hyperfine states $\ket{F=1,m_F=-1}$ and $\ket{F=2,m_F=-2}$ of odd and even sites are split by $28$ kHz. Atoms on the odd or even sites can then be adiabatically transferred to $\ket{F=2,m_F=-2}$, which is resonant with the imaging laser. The initial state with $\langle \hat{n}_\text{matter}\rangle = 1$ and $\langle \hat{n}_\text{gauge}\rangle = 0$ is prepared by simply transferring and removing all atoms on odd sites with the imaging laser. On the other hand, for detecting atom number density, we transfer and then extract $\langle \hat{n}_\text{gauge}\rangle$ and $\langle \hat{n}_\text{matter}\rangle$ with in-situ absorption imaging of odd and even sites successively in one experimental sequence. 

To access various gauge-invariant initial states with $\langle \hat{n}_\text{matter}\rangle < 1$ in \Fig{finitialcond}, we start with the $\langle \hat{n}_\text{matter}\rangle = 1$ state, and using an optimized version of the adiabatic ramp described in \cite{Yang2020}, we can prepare states ranging from $\langle \hat{n}_\text{matter}\rangle = 1$ to $\langle \hat{n}_\text{matter}\rangle = 0.21$ by stopping the ramp at certain time $\tau$ (see \Fig{finitialcond}\textbf{b}).

To obtain the gauge violation $\eta$ during the quench dynamics, we measure the density of odd-number of atoms occupying the gauge sites, which is done by parity projection with the photoassociation (PA) laser. The PA laser excites pairs of atoms to molecular states which quickly decay to free channels by emitting photons, thereby imparting atoms with the kinetic energy required to escape from the lattice potential. After a certain evolution time, a $20$ ms PA laser pulse was applied with the intensity of $0.67$ $\text{W/cm}^2$. An efficiency of $98\%$ was reached for removing doublons.

\subsection{Identifying the resonance condition} 

\begin{figure}
	\centering
	\includegraphics[width=83mm]{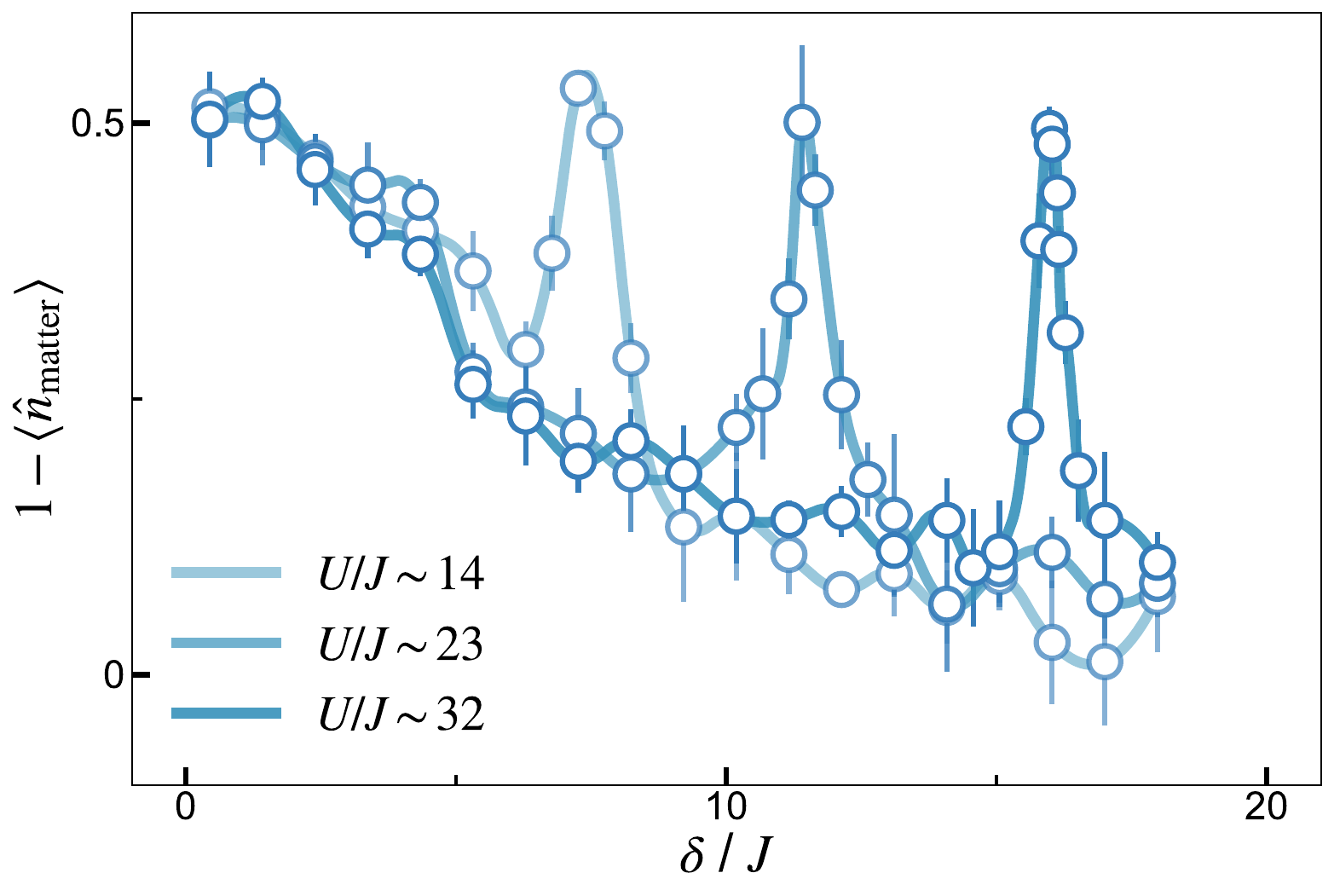}
	\caption{Late-time parameter space. The average atom population on gauge sites after $120$ ms evolution, scanned over the parameters $\delta/J$. The peaks in the figure correspond to the resonance condition of the correlated second-order hopping, $\delta = U/2$.}\label{fdeltascan}
\end{figure}
We first perform a 2D parameter scan to find the resonance condition of correlated second-order tunneling that realizes the gauge theory (see \Fig{ffreq}\textbf{c}). Starting from the initial state with $\langle \hat{n}_\text{matter}\rangle = 1$, which is the ground state of Hamiltonian Eq.~\eqref{eq:QLM} at $m\to-\infty$, we bring the system out-of-equilibrium by turning on the gauge coupling $\kappa$ while quenching the effective mass from $m\rightarrow-\infty$ to $m\approx0$ with $m=\delta-U/2$. By setting the relative phase $\varphi=\pi/4$, we configure the superlattice in the form of a \textit{staggered} potential (see \Fig{fmap}\textbf{b}), where $\delta$ is mainly determined by the long lattice depth while $J$ is set by the short lattice depth. The on-site interaction $U$ is then determined by the trapping frequency of the short lattice along the $x$, $y$, and $z$ axes. Each data point in the parameter scan is obtained with $120$ ms of quench evolution during which the system roughly relaxes to a steady state. The scans are done by fixing $U$ and $J$ while changing $\delta$ for each quench evolution, where \Fig{fdeltascan} shows three examples of the scans. The suppression of direct tunneling can be seen as $\langle \hat{n}_\text{matter}\rangle$ goes to 1 with increasing $\delta/J$. At $\delta \approx U/2$, a resonance peak can be observed corresponding to $m \approx 0$ in the gauge theory. We perform our further studies in the vicinity of these resonance peaks.

\subsection{Approaching the gauge theory} 

\begin{figure}
	\centering
	\includegraphics[width=83mm]{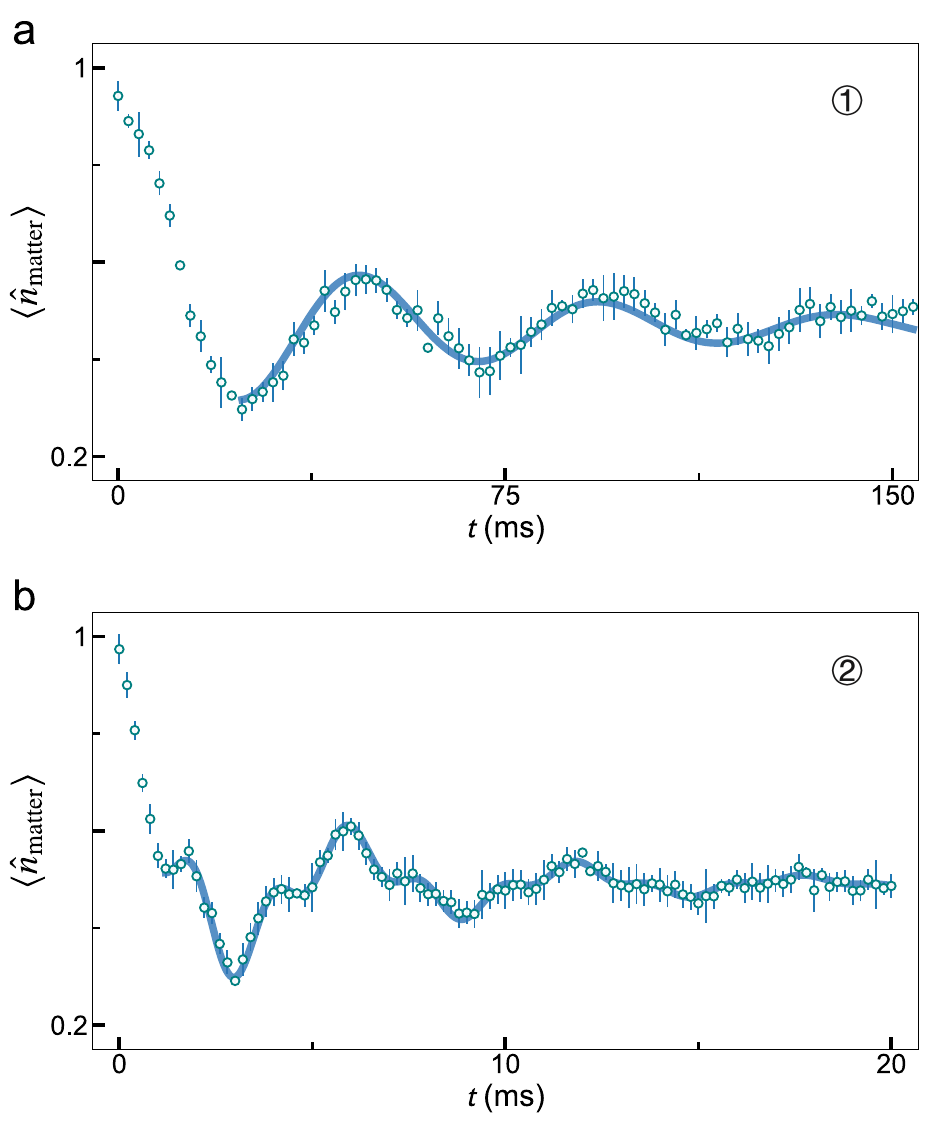}
	\caption{Damped-sine fitting for different staggering. (a) We extract the oscillation frequency ($f_\text{exp} = 21$ Hz) at large staggering $\delta/J = 16$. (b) At staggering $\delta/J = 6.5$, two oscillation frequencies ($f_\text{1} = 507$ Hz, $f_\text{2} = 172$ Hz) are extracted from the data. The solid blue curves result from the damped-sine fitting with Eq.~\eqref{eq:dual-sine}. They extend over the times considered for the fit.}\label{fdamp-sine}
\end{figure}

For cases in \Fig{ffreq}\textbf{b} where we have the strongest quench from $m\rightarrow -\infty$ to $m=\delta-U/2=0$, the many-body oscillation frequency of the gauge theory solely depends on the coupling strength $\kappa$. We associate the correlated second-order tunneling $8\sqrt{2}J^2/U$ in the Bose--Hubbard simulator with $\kappa$ through perturbation theory, and gauge violations introduced by direct tunneling $J$ leads the Bose--Hubbard simulator to deviate from the gauge theory. With large $U/J$, gauge violation is suppressed and the frequency of the Bose--Hubbard simulator approaches the gauge theory. To quantitatively identify this approach to the gauge theory, we extract the oscillation frequency $f_\text{exp}$ in experiment with a damped-sine fitting, and calculate its ratio with the gauge-theory frequency $f_\text{gauge}$. And to capture both first and second-order tunneling in the Bose--Hubbard simulator, we use the dual-frequency damped sine-wave function for the fitting:

\begin{equation}
\label{eq:dual-sine}
\ n = n_0 + A_1e^{-\gamma_1 t}\text{sin}(2\pi f_1 t+b_1) + A_2e^{-\gamma_2 t}\text{sin}(2\pi f_2 t+b_2).
\end{equation}
where $\omega_{1,2}$ represent the two angular frequencies extracted from the oscillations and $\gamma_{1(2)}$ indicates the damping rate for each frequency. For quench dynamics in the gauge-invariant regime ($\delta/J>10$, case 1 in \Fig{ffreq}\textbf{a}) and the ``nongauge'' regime ($\delta/J<2$, case 3 in \Fig{ffreq}\textbf{a}), oscillations are dominated by one of the tunneling processes and fitting gives identical $f_{1}$ and $f_{2}$, whence we take the average $f_\text{exp} ={(f_1+f_2)/2}$. However, in the ``cross-over'' regime ($2<\delta/J<10$, case 2 in \Fig{ffreq}\textbf{a}), a clear beating of two frequencies can be seen, and both frequencies are extracted. As the higher frequency comes from direct tunneling $J$, we take the slower frequency which captures the second-order tunneling $f_\text{exp} ={\min(f_1,f_2)}$. To avoid turning-on effects, we start the fit from the first minimum, see \Fig{fdamp-sine}.

The numerical data shown in \Fig{ffreq}\textbf{b} (solid curve) is obtained from $t$-DMRG with $L = 32$ sites. We calculated 129 sets of $t$-DMRG data with increasing staggering at different conditions ($\delta-U/2= 0$, $\delta-U/2=\pm3$ Hz) and applied the damped sine-wave function on the data. The solid curve in \Fig{ffreq}\textbf{b} represents the result at the resonance condition ($\delta-U/2= 0$) while the grey area describes the boundary of off-resonance conditions ($\delta-U/2= \pm3$ Hz).

\subsection{Quantifying the gauge violation}

\begin{figure}
	\centering
	\includegraphics[width=83mm]{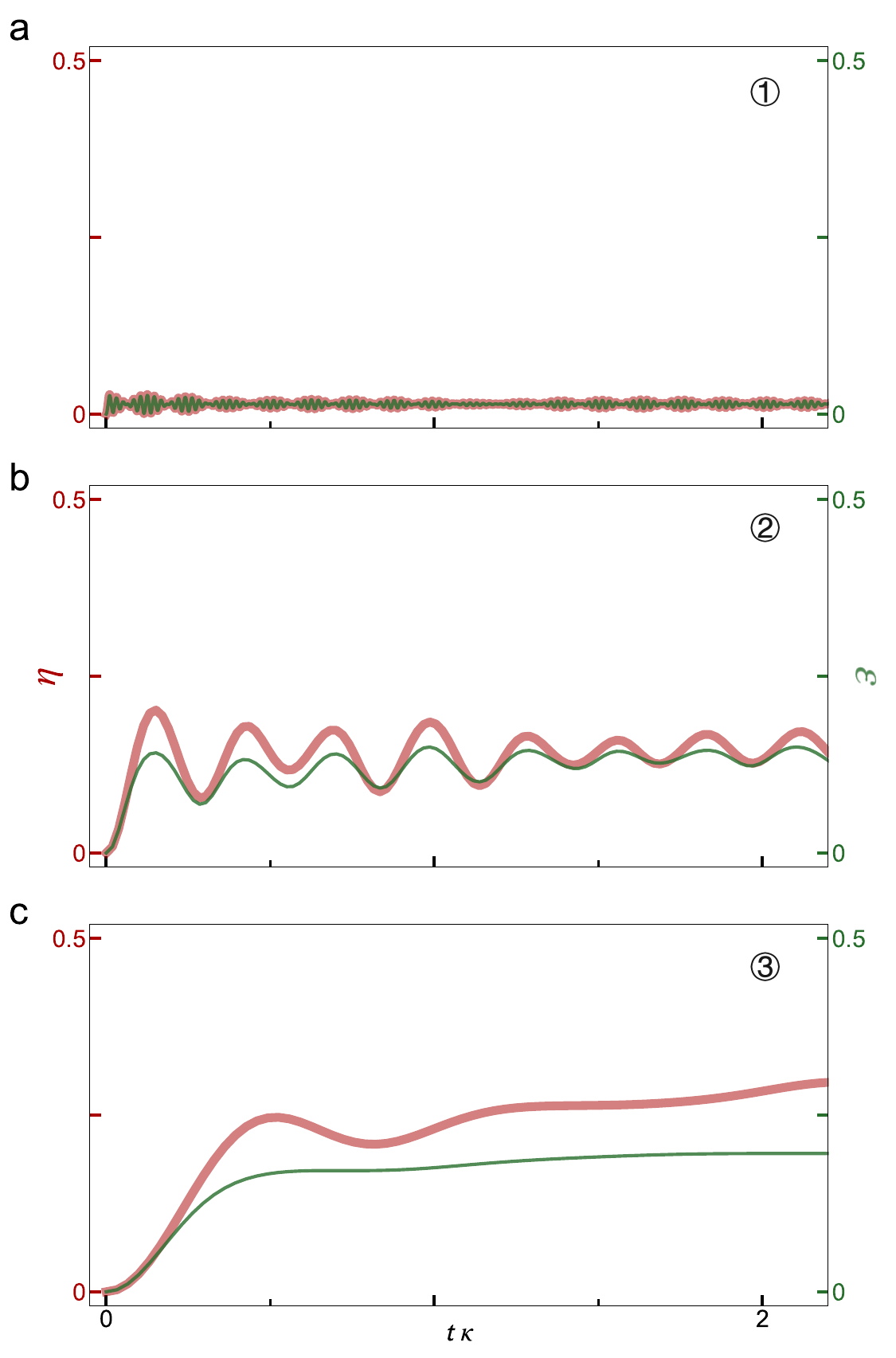}
	\caption{\textbf{Gauge violation measure.} We numerically compare our experimentally accessible local violation measure $\eta$ (thick red line) to a global measure $\varepsilon$ (thin green line, see text). We find reasonable agreement for both the regime close to (case 1) and far away from the gauge theory (cases 2 and 3). In the second, non-gauge regime, the system quickly reaches the maximum global violation $1/L_\text{g}$. Here, the system size is $L_\text{g} = 5$.}\label{fvio}
\end{figure}

Overall, our results show a faithful implementation of the gauge theory for the considered quench dynamics starting from far-from-equilibrium initial states. We achieve this by strongly suppressing processes that violate the gauge symmetry. To further quantify this, in the following we consider the direct tunneling for the quench dynamics starting from the "fully filled" initial state. In \Fig{fvio}, we show the average gauge-violation per site $\eta$, as defined in the main text, and compare it to the global measure $\varepsilon = (1- \langle \hat{P}_G \rangle) /L_\text{g}$, where $\hat{P}_G$ is the projector onto the exact gauge-invariant subsector of the full Hilbert space. 

In panel \textbf{a}, we see that for the employed evolution parameters and initial state, the two measures agree in the gauge regime (case 1). Close to the gauge regime, the quantitative agreement can be understood from a simple error model assuming that gauge errors appear at every site independently and at the same rate. Suppose this rate is given by our experimentally accessible measure $\eta$, then the probability of remaining in the gauge-invariant subsector is approximately given by $\langle\hat{P}_G \rangle \sim (1-\eta)^{L_\text{g}} \approx 1- L_\text{g} \eta$, such that $\varepsilon \sim \eta$. While this is strictly only valid for~$L_\text{g} \eta \ll 1$, our numerical results in \Fig{fvio} suggest reasonable agreement even away from the gauge-invariant regime (cases 2 and 3 in panels \textbf{b} and \textbf{c}).

\subsection{Thermal ensembles for the $\mathrm{U}(1)$ quantum link model}
For the calculation of the thermal values for the steady states in our quench dynamics, we utilize the diagonal, canonical, and microcanonical ensembles. When the long-time average of the unitary dynamics agrees with the expectation value predicted by the diagonal ensemble, this indicates that the system has equilibrated. Furthermore, when this expectation value additionally agrees with those obtained from the microcanonical and canonical ensembles, this implies thermalization \cite{rigol2008thermalization}.

In exact diagonalization, we work in an effective subspace of the total Hilbert space of the $\mathrm{U}(1)$ quantum link model, such that only gauge-invariant states within the physical sector are accounted for. We employ open boundary conditions, even though at sufficiently large system sizes, there is little difference from the case with periodic boundary conditions.

\subsubsection{Diagonal ensemble}
Quenching the initial state $\ket{\Psi_0}$ with the $\mathrm{U}(1)$ quantum link model Hamiltonian $\hat{H}$ leads to the dynamics
\begin{align}\nonumber
    \hat{\rho}(t)=&\,e^{-i\hat{H}t}\ket{\Psi_0}\bra{\Psi_0}e^{i\hat{H}t}\\\nonumber
    =&\,\sum_{m,n}e^{-i(E_m-E_n)t}\bra{E_m}\ket{\Psi_0}\bra{\Psi_0}\ket{E_n}\ket{E_m}\bra{E_n}\\\nonumber
    =&\,\sum_{m\neq n}e^{-i(E_m-E_n)t}\bra{E_m}\ket{\Psi_0}\bra{\Psi_0}\ket{E_n}\ket{E_m}\bra{E_n}\\\label{eq:rho}
    &+\sum_{m}\big\lvert\bra{E_m}\ket{\Psi_0}\big\rvert^2\ket{E_m}\bra{E_m},
\end{align}
where $\{\ket{E_m}\}$ is the eigenbasis of $\hat{H}$. The long-time average of any observable will be dominated by the second term of the last equality in Eq.~\eqref{eq:rho}. This gives rise to the diagonal ensemble
\begin{align}
    \hat{\rho}_\text{DE}=\sum_{m}\big\lvert\bra{E_m}\ket{\Psi_0}\big\rvert^2\ket{E_m}\bra{E_m},
\end{align}
which describes the steady state to which the system equilibrates at late times.

\subsubsection{Canonical ensemble}
Quenching the initial state $\ket{\Psi_0}$ with the $\mathrm{U}(1)$ quantum link model Hamiltonian $\hat{H}$ gives rise to the energy 
\begin{align}
    E_0=\bra{\Psi_0}\hat{H}\ket{\Psi_0}.
\end{align}
This is a conserved quantity throughout the whole unitary time evolution of our isolated system. As such, if there is a canonical ensemble,
\begin{align}
\hat{\rho}_\text{CE}=\frac{e^{-\beta \hat{H}}}{\Tr{e^{-\beta \hat{H}}}},  
\end{align}
to which the system thermalizes, then it must satisfy 
\begin{align}
    E_0=\Tr{\hat{H}\hat{\rho}_\text{CE}}.
\end{align}
This is a single equation with the only unknown being the inverse temperature $\beta$ of the thermal steady state, and it can be numerically solved using, e.g., Newton's method. If thermalization does occur, then the temporal average of the unitary dynamics of any local observable $\hat{O}$ over sufficiently long evolution times should be equivalent to the prediction by the canonical ensemble,
\begin{align}\label{eq:CE}
    \lim_{t\to\infty}\frac{1}{t}\int_0^t ds\,\bra{\Psi_0}e^{i\hat{H}s}\hat{O}e^{-i\hat{H}s}\ket{\Psi_0}\equiv\Tr{\hat{O}\hat{\rho}_\text{CE}}.
\end{align}
Moreover, in case of thermalization, then the system must have also equilibrated, meaning additionally to Eq.~\eqref{eq:CE} that
\begin{align}
    \Tr{\hat{O}\hat{\rho}_\text{CE}}\equiv\Tr{\hat{O}\hat{\rho}_\text{DE}},
\end{align}
i.e., that the expectation value of a local observable should be the same when obtained through the canonical or diagonal ensemble.

\begin{figure*}
	\centering
	\includegraphics[width=140mm]{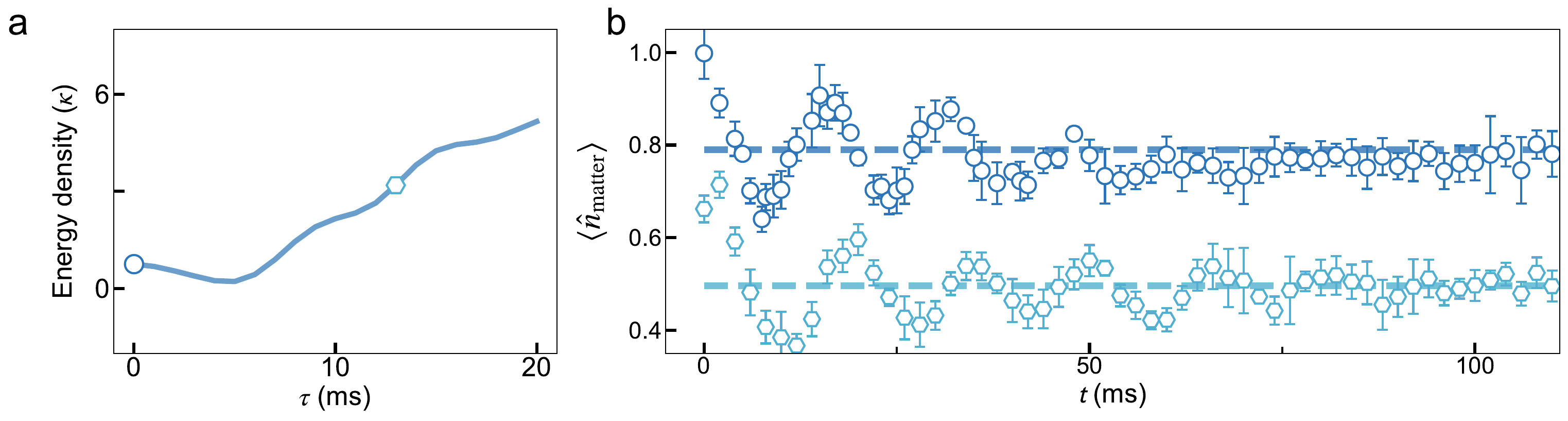}
	\caption{Nonequilibrium thermalization dynamics for quench parameter $m=-0.8\kappa$. (a) Initial-state energy densities for $m=-0.8\kappa$. We choose two initial states with unequal energy density to perform the dynamics. (b) Here we show the nonequilibrium thermalization dynamics for the initial states shown in panel \textbf{a}, and without the same energy density, they do not converge to a common steady state. The dashed line in panel \textbf{b} are thermal values calculated from the gauge theory thermal ensemble with corresponding energies. }\label{fthermal2}
\end{figure*}

\subsubsection{Microcanonical ensemble}
The microcanonical ensemble is constructed from the eigenstates of the quench Hamiltonian $\hat{H}$ that lie within the energy shell $[E_0-\Delta E,E_0+\Delta E]$. Formally, this ensemble is
\begin{align}
    \hat{\rho}_\text{ME}=\frac{1}{\mathcal{N}_{E_0,\Delta E}}\sum_{m;\,\lvert E_m-E_0\rvert\leq\Delta E}\ket{E_m}\bra{E_m},
\end{align}
where $\mathcal{N}_{E_0,\Delta E}$ is the number of eigenstates in the energy shell $[E_0-\Delta E,E_0+\Delta E]$. We have chosen the half-width to be $\Delta E=0.05\kappa$, although the thermal value predicted by the microcanonical ensemble is robust to the exact value of $\Delta E$.

In case of thermalization, different thermal ensembles will give the same expectation value for a given observable $\hat{O}$. Given that thermalization also indicates equilibration, this leads to
\begin{align}\label{eq:finale}
    \Tr{\hat{O}\hat{\rho}_\text{ME}}\equiv\Tr{\hat{O}\hat{\rho}_\text{CE}}\equiv\Tr{\hat{O}\hat{\rho}_\text{DE}}.
\end{align}

\subsubsection{Infinite-temperature state}
We consider the fully filled initial state with $\langle \hat{n}_\mathrm{matter}\rangle = 1$ in the gauge theory, which corresponds to $\ket{\psi_0} = \ket{...10101...}$ as depicted in the first line of \Fig{app:Gauss-law}. For $m=0$, we have
\begin{align}
	E_0 = \bra{\psi_0} \hat{H}_\mathrm{gauge} \ket{\psi_0} = 0 \; .
\end{align}
The operator
\begin{align}
	\hat{V} = \prod_{\ell} \hat{S}^z_{\ell,\ell+1} 
\end{align} 
anti-commutes with the Hamiltonian, $\{\hat{V},\hat{H}_\mathrm{gauge}\}=0$, such that for any eigenstate $\ket{e}$ with $\hat{H}_\mathrm{gauge}\ket{e} = E\ket{e}$ there is an associated state $\ket{\tilde{e}}= \hat{V}\ket{e}$ with eigenvalue $\hat{H}_\mathrm{gauge}\ket{\tilde{e}} =- E\ket{\tilde{e}}$. Hence, the spectrum is symmetric around $E_0 = 0$. Since $\ket{\psi_0}$ is an eigenstate of $\hat{V}$ with eigenvalue $\pm1$ it is symmetrically distributed in the spectrum which is in conflict with a thermal state except for $T=\infty$. 


\subsection{Sensitivity of thermal value to the conserved quantity}
The thermal value of a thermal ensemble is determined by its conserved quantities. In \Fig{finitialcond}\textbf{d}, we show that starting in initial states with the same energy density leads to quench dynamics that converge to steady states with the same thermal value for quench parameters $m=0$ and $m=-0.8\kappa$. To further investigate the role of the energy density in thermalization dynamics, we choose two initial states with unequal energy density for quench parameter $m=-0.8\kappa$, as shown in \Fig{fthermal2}\textbf{a}. We apply the preparation ramp to initialize the states with different energy densities at ramp times $\tau = 0$ and $13$ ms. In \Fig{fthermal2}\textbf{b}, the quench dynamics due to these two initial states show a clear deviation from each other in their long-time steady states, which agrees well with our predictions.

\end{document}